\documentstyle[aasms4]{article}
 
\begin{document}
 
\title{RXTE Discovery of Coherent Millisecond Pulsations during an X-ray Burst from KS 1731-260}
\author{Donald A. Smith\altaffilmark{1}, Edward H. 
Morgan\altaffilmark{1}, Hale Bradt\altaffilmark{1}}
\altaffiltext{1}{Center for Space Research and Department of Physics, MIT, Cambridge, MA, 02139, USA}
\authoremail{dasmith@space.mit.edu}
 
\begin{abstract}
A highly coherent $523.92\pm0.05$ Hz periodic X-ray signal has been
observed during a type I X-ray burst from the low-mass X-ray binary
system KS 1731-260 with the PCA on RXTE.  The spectral evolution of
the burst indicates photospheric-radius expansion and contraction.
The 524 Hz signal occurred at the end of the contraction phase, lasted
for $\sim$2 s, was highly coherent (Q $\gtrsim$ 900), and had a pulse
fraction (ratio of sinusoidal amplitude to mean count rate) of
$6.2\pm0.6$\%.  KS 1731--260 is one of only three systems that have
exhibited high-coherence millisecond oscillations during X-ray bursts
and the first reported where the pulsations are associated with
photospheric contraction.  These coherent signals may be interpreted
as a direct indication of the neutron star spin.
\end{abstract}


\section{Introduction and Observations}

Accreting low-mass X-ray binaries (LMXBs) may be the progenitors of
the millisecond radio pulsars.  The long life of these systems, during
which angular momentum is transferred to the neutron star through
accretion, is thought to lead to the millisecond spin periods observed
in the fastest radio pulsars \markcite{smarr1976}(Smarr \& Blandford
1976; \markcite{bhatt1995} Bhattacharya 1995).  Prior to the launch of
the {\it Rossi X-ray Timing Explorer} ({\it RXTE}), however, no
convincing spin periods of less than 131 ms (in Aql X-1;
\markcite{schol1991} Schoelkopf \& Kelley 1991) had been found in any
LMXB.  High time resolution and large effective area give {\it RXTE}
sensitivity to pulse frequencies into the kilohertz range, which
allows users to search for coherent signals that might indicate
neutron stars spinning with millisecond periods.  Three examples of
such signals have been reported: a 363 Hz (2.75 ms) oscillation in six
bursts from 4U 1728--34 \markcite{stro1996a} (Strohmayer et al.
1996a), a 524 Hz (1.91 ms) oscillation in a single burst from KS
1731--260 \markcite{morga1996} (Morgan \& Smith 1996), and a 589 Hz
(1.70 ms) oscillation in three bursts from an unidentified source near
GRO J1744--28 \markcite{stro1996b} (Strohmayer, Lee \& Jahoda 1996b).
Here we provide a detailed report of the KS 1731--260 observation.

The galactic x-ray source KS 1731--260 was first discovered in
outburst during August 1989 with the imaging spectrometer TTM on the
{\it MIR-KVANT} observatory \markcite{sunya1989} (Sunyaev 1989).  It
was observed for twenty-four 1000-s intervals over the course of 15
days, and its intensity ranged from 50 to 100 mCrab in the 2--27 keV
band.  The spectrum of the persistent emission was found to be
well-fit by a thermal bremsstrahlung model with a temperature of
$5.7\pm0.3$ keV.  Three type I X-ray bursts were observed from KS
1731--260 during this period.  The bursts lasted 10--20 seconds each,
and reached intensities up to 0.6 Crab \markcite{suny1990b} (Sunyaev
et al. 1990b).  The presence of type I bursts identified the compact
object as an accreting neutron star.  No pulsations were detected.  In
short, the source exhibited characteristics typical of a low-mass
X-ray binary system \markcite{barre1992} (Barret et al.  1992).

Subsequent detections of KS 1731--260 were sporadic, which implied that
the source is transient in nature.  In 1990, the source was detected
on two occasions (April 4 and August 23) by the ART-P imaging
instrument aboard the {\it GRANAT} spacecraft \markcite{suny1990a}
(Sunyaev et al. 1990a, \markcite{suny1990b} b).  In 1990 and 1991, the
SIGMA telescope detected the source only once (on 1991 March 14 during
a $\sim$20 hr exposure) in a series of twenty observations of its
location \markcite{barre1992} (Barret et al.  1992).  It was also
detected during the {\it ROSAT} all-sky survey \markcite{prede1995}
(Predehl \& Schmitt 1995).

Results from the {\it RXTE} All-Sky Monitor (ASM) reveal that KS
1731--260 is currently active.  The source was detectable from the
time of ASM turn-on in January 1996 through November 1996 with a
typical intensity of $\sim$160 mCrab ($\sim$1.5-12 keV).  Further
information about the ASM and its observations can be found in
\markcite{levin1996} Levine et al.  (1996) and also on the World Wide
Web at http://space.mit.edu/XTE/XTE.html.
 
The ASM reports led us to request public observations of KS 1731--260
with {\it RXTE}'s Proportional Counter Array (PCA).  Three
observations with all five Proportional Counter Units (PCUs) were
carried out for a total on-source time of 23400 s, divided into ten
uninterrupted intervals (See Table 1). The Experiment Data System
(EDS), the on-board computer that processes the data from the PCA, was
configured to provide a primary data mode with 32 energy channels
across the PCA's energy range of $\sim$2--90 keV and 62 $\mu$s
($2^{-14}$ s) time resolution.  This event mode will only record the
first $\sim$16,000 photons in any given second, but even at higher
count rates, the event-mode data in a given second still provides
useful rates.  The arrival time of the last recorded photon is treated
as the beginning of a data gap that ends with the beginning of the
next second.  To recover information lost in such a gap, we ran burst
trigger and catcher modes in parallel with the event mode.  The chosen
burst catcher mode has 125 $\mu$s time resolution but no energy
resolution.

\section{Analysis and Results}

During the ten observing intervals, the total count rate of `good'
events in the PCA from the typical source emission and background
varied between 1619 and 3459 c/s.  The typical background rate was
$\sim$130 c/s.  The measured intensity above 27 keV was consistent
with background during all observations.  A single X-ray burst was
observed during this program with onset at 1996 July 14 04:16:17.5
(UTC).  The burst is displayed in Figure 1 in three different photon
energy bands, which were selected such that each band has roughly the
same count rate during the pre-burst emission.  The spectrum of the
emission immediately prior to the burst is well-fit (reduced $\chi^{2}
= 1.0$ for a 60-s exposure of one layer of one PCU) by an absorbed
thermal bremsstrahlung model with a temperature of $kT = 5.5\pm0.2$
keV and a hydrogen column density of $(60\pm3)\times10^{21}$ ${\rm
cm}^{-2}$.  There is some evidence that the source exhibits atoll-like
behavior.  A color-color diagram of these observations shows the
source climbing through a banana-like state as its intensity increases
on July 14, but staying in a soft island-like state at the lower count
rates during the two August observations.

The persistent emission data (2--27 keV) were searched for
periodicities using a standard fast Fourier transform (FFT) algorithm.
The time series were corrected for the difference in photon arrival
time between Earth and the solar system barycenter with the Jet
Propulsion Laboratory DE-200 solar system ephemeris
\markcite{stand1992} (Standish et al.  1992).  The 62-$\mu$s event-
mode data were transformed in 64-s segments and averaged together with
appropriate weighting.  The power of the Poisson noise was normalized
to 2 in the power density spectra (PDS) \markcite{leahy1983} (Leahy et
al. 1983).  The observations on July 14 showed very low-frequency
noise with a power law index of --$1.00\pm0.06$, consistent with
expectations for an atoll source in the banana state, but there was no
strong high-frequency noise component present in the August
observations, as would be expected from an atoll source in its island
state, as suggested by \markcite{hasin1989} Hasinger \& van der Klis
(1989).  No coherent periodic signals are detected in the PDS to a
$2\sigma$ upper limit of 0.09\% on the pulse fraction.  The pulse
fraction is defined as the ratio of the amplitude of an assumed
sinusoidal pulse profile to the mean count rate.  All upper limits to
pulse fractions are calculated with the algorithm described by
\markcite{vaugh1994} Vaughn et al.  (1994).

There are no visible quasi-periodic oscillations (QPO).  In order to
indicate an upper limit on the strength of an undetected QPO peak, we
fit a Lorentzian profile to the PDS, adopting the 64.0 Hz width of the
$\sim$1 kHz QPO peak observed from 4U 1728--34 by \markcite{stro1996a}
Strohmayer et al.  (1996a) The centroid was positioned at the bin with
the maximum power near 1 kHz.  The resulting best-fit normalization
indicates that if a QPO of 64.0 Hz width were in the non-burst
emission from KS 1731--260, its integrated rms power would be less
than 1\%.

The burst was also searched for periodicities.  A transform of the
0.5-s rise data produced no coherent features in any bin to a
$2\sigma$ upper limit on the pulse fraction of 12.6\%.  The next ten
seconds of the burst were divided into consecutive, adjacent 1-s
intervals and transformed separately.  A significant peak appears at
524 Hz in two PDS from adjacent intervals, at power levels of 29.75
and 33.44 (Fig. 2).  This 2-s interval is indicated with vertical
dashed lines in Figure 1.  A transform of this interval increases
frequency resolution to 0.5 Hz.  The best-fit Gaussian function to
this peak has a full-width half-maximum of $0.54^{+0.05}_{-0.27}$ Hz.
The lower limit of the coherence of the signal is therefore Q $\gtrsim
900$, where Q is defined as the ratio of the peak center to the peak
width.  There is no evidence for any other periodicities in the
frequency range of 15-8192 Hz (including harmonics of the 524 Hz
signal) to $2\sigma$ upper limits on the pulse fraction at any
frequency of 4.4\%.  There is no evidence for a signal at 524 Hz
during the burst rise to a $2\sigma$ upper limit of 4.9\%.  Power
density spectra for 8 adjacent 2-s intervals that begin after the
burst rise is over (excluding the 2-s interval when the 524 Hz signal
is observed) place $2\sigma$ upper limits on the pulse fraction at any
frequency that increase from 4.4\% to 12.9\% with decreasing count
rate, while the $2\sigma$ upper limit on the pulse fraction at 524 Hz
varies between 1.5\% and 5.0\%.

It is highly improbable that the peaks at 524 Hz displayed in Figure
2 are due to random noise alone.  The probability of a high
(Power $\equiv P_{0} \geq 29.75$) peak occurring at random in any
one of the $N_{\rm tot} = 81920$ independent frequency bins from the
ten 1-s intervals is
\[{\rm prob}(P_{\rm noise} > P_{0}) = 1 - (1 - e^{\slantfrac{-P_{0}}{2}})^{N_{\rm tot}} = 1 - (1 - e^{-14.875})^{81920} = 0.0280\]
(Vaughn et al. 1994). The probability of a second peak at $P_{0}
\geq 33.44$ appearing at random in the same frequency bin of an
adjacent, independent PDS is only $1 - (1 - e^{-16.72})^{2} = 1.1
\times 10^{-7}$.  In the PDS from the 2-s interval, the strength of
the feature increases to 54.25.  The probability for a peak of this
height or higher to appear at random in any bin of this PDS is $2.7
\times 10^{-8}$.

The frequency of the pulsations was constrained by means of $\chi^{2}$
sinusoidal fits to the counts from the burst-catcher mode, which had
no gaps at this high count rate.  We split the 2-s interval into 0.5-s
segments and found that the frequency is constant with a 1$\sigma$
upper limit on its rate of change of 0.5 Hz/s.  For the entire 2-s
interval, the best-fit constant frequency is $523.92\pm0.05$ Hz.  The
evolution of the pulsed amplitude was obtained from fits to 0.25-s
intervals; it rose above a 1$\sigma$ detection at 04:16:20.0$\pm$0.25
s and fell below that level again at 04:16:22.0$\pm$0.5 s.

The energy dependence of the 524 Hz signal was obtained from
sinusoidal fits to the event-mode data.  The signal was strongest at
higher energies (Fig. 3).  At 7--27 keV, the pulse fraction is
$9.0\pm0.8$\%.  At 2--7 keV, a 2$\sigma$ upper limit of 4.1\% was
obtained via the algorithm described in \markcite{chakr1995}
Chakrabarty et al.  (1995).

The spectral evolution of the burst was studied through fits to the
event-mode data in 0.25-s intervals over the course of the burst.
Model spectra were folded through a PCA response matrix from a
pre-release version of FTOOLS 3.6.1
(http://heasarc.gsfc.nasa.gov/docs/xte/whatsnew/software.html).  A
spectrum from a 1000-s interval immediately prior to the burst was
subtracted from each of these spectra before any model fitting.  This
subtraction assumes that the mechanism that produces the pre-burst
emission remains unaffected by the processes of the burst, which may
not reflect the physics of the neutron star environment.

The best values of the reduced-$\chi^{2}$ statistic were achieved with
a Planckian blackbody spectrum.  The value for the column density
$N_{\rm H}$ was fixed at the {\it ROSAT} value of
$(10.0\pm1.9)\times10^{21}$ ${\rm cm}^{-2}$ \markcite{prede1995}
(Predehl \& Schmitt 1995). The results of these fits are displayed in
Figure 4.  The somewhat high values of the reduced $\chi^{2}$
statistic appear to arise from systematic deviations of order
$\sim$20\% in the burst spectra from the Planckian model.  Attempts to
account for these deviations by adding a second continuum component
were unsuccessful.  A deviation in the form of a spectral hardening
due to the dominance of of electron scattering is expected at high
luminosities \markcite{casto1974} (Castor 1974; \markcite{londo1984}
London, Tamm, \& Howard 1984) and has been observed in bursts from
other sources (e.g. \markcite{vanpa1990} van Paradijs et al. 1990).
Our deviations show more structure than can be explained by a simple
temperature shift, but in view of the relatively poor spectral
resolution of this data mode it seemed unproductive to apply
multi-component models with narrow spectral features to these
deviations.  Therefore the radii and temperatures given in the figure
must be considered uncertain in a systematic sense and treated with
caution.

Spectral fits to other simple models such as thermal Bremsstrahlung or
a power law give reduced $\chi^{2}$ values higher by factors up to 32.
Allowing $N_{\rm H}$ to float improved the Bremsstrahlung fit to the
same level as the blackbody fit at the cost of increasing $N_{\rm
H}$ to $(95\pm2)\times10^{21}$ ${\rm cm}^{-2}$, which is inconsistent
with the ROSAT measurements.

The best-fit parameters from the Planckian models exhibit behavior
expected from a radius-expansion burst (Fig. 4).  Radius-expansion
bursts are believed to occur when the luminosity of the burst reaches
the Eddington limit, and the atmosphere of the neutron star
temporarily expands due to radiation pressure \markcite{lewin1984}
(Lewin, Vacca, \& Basinska 1984; \markcite{tawar1984} Tawara et al.
1984).  In this case, the $\sim$0.75-s expansion phase is indicated by
the increase of the normalization of the model (which represents the
square of the blackbody radius at 10 kpc) and the corresponding
decrease in the temperature.  During the subsequent $\sim$1.5-s
contraction, the temperature increases as the atmosphere settles back
to the surface.  The temperature reaches a maximum as the radius
reaches a minimum.  Thereafter the radius remains constant as the
temperature decreases.  The 2.0-s period when the 524 Hz oscillations
appear are again marked by vertical dashed lines.

The best-fit spectral temperature is expected to be overestimated by a
factor of approximately 1.4 at near-Eddington luminosities due to the
reduction of the radiation source function through electron scattering
\markcite{londo1984} (London, Tamm, \& Howard 1984).  The maximum
effective temperature during this burst then becomes $1.7\pm0.2$ keV.
\markcite{marsh1982} Marshall (1982) calculates that blackbody
temperatures higher then 1.96 keV (for a helium-rich atmophere; 1.67
keV for a hydrogen atmosphere) contradict the mass-radius relations
derived from both soft and hard equations of state
\markcite{arnet1977} (Arnett and Bowers 1977) and are therefore
impossible for a isotropically-radiating neutron star to attain.  The
spectral hardening correction brings our maximum temperature into
agreement with these limits.

The evolution of the blackbody luminosity follows from the temperature
and the radius under the assumption of spherical symmetry (Fig. 4c).
The luminosity during the expansion and contraction phases of the
burst is consistent with a constant, as is expected if the burst is
Eddington-limited.  Therefore a rough estimate of the distance to the
source can be calculated.  For a 1.4 M$_\odot$ neutron star with a
helium-rich envelope, the Eddington luminosity as observed from a
great distance is approximately $2.7 \times 10^{31}$ W (see e.g.,
\markcite{haber1987} Haberl et al. 1987).  The blackbody fits then
yield a distance of $8.8\pm0.3$ kpc.  At this distance, the
``touchdown radius'' indicated by the value of the normalization at
the end of the contraction period is $9\pm1$ km.  The spectral
hardening at high luminosities would imply this result is
underestimated by a factor of order 2.  The errors are formal 90\%
confidence limits that include a 10\% error estimate for the Eddington
luminosity added in quadrature.

\section{Discussion}

We believe that rotational modulation of asymmetric emission is the
simplest and most plausible mechanism for producing the observed
524 Hz oscillations.  The high coherence (Q $\gtrsim$ 900) of the peak
argues for a highly coherent mechanism such as stellar rotation.  The
resulting spin period of 1.91 ms is consistent with predictions of
spin periods in LMXBs \markcite{smarr1976} (Smarr \& Blandford 1976;
\markcite{verbu1995} Verbunt \& van den Heuvel 1995) as well as the
observed periods of millisecond ``recycled'' radio pulsars
\markcite{bhatt1995} (e.g. Bhattacharya 1995; \markcite{backe1982}
Backer et al. 1982; \markcite{fruch1988} Fruchter, Stinebring \&
Taylor 1988).

A key characteristic of the oscillations that any suggested model must
explain is the point in the evolution of the burst at which they
appear.  No signal is seen during the rise of the burst, as would be
expected if the asymmetry that gives rise to the modulation were due
to a slowly propagating burning front as suggested by
\markcite{bilds1995} Bildsten (1995).  Rather, the pulsations appear a
full two seconds after the burning is over, at the end of the
contraction phase, and they persist into the cooling phase.  It is
difficult to determine exactly when the pulsations vanish, as the
detection threshold increases with decreasing count rate.  All that
can be said for certain is that they vanish during the cooling phase,
for the $2\sigma$ upper limit on their pulse fraction in the
persistent emission is 0.07\% of the mean count rate.
  
It has been suggested (Marshall 1997, in preparation) that the
presence of a weak magnetic field could cause an anisotropy in the
atmospheric opacity.  Emission modulation due to this anisotropy
would be visible if the atmospheric scale height were small, a
vertical temperature gradient were present, and the magnetic field
axis were not aligned with the rotation axis.  We hypothesize that
early in the burst, the atmosphere expands to a height where the
magnetic field is too weak to cause a significant opacity difference.
As the energy from the burst is radiated away in the cooling phase,
the atmosphere returns to its isothermal quiescent state, which would
explain the disappearance of the pulsations in the burst as well as
their absence at other times.

Observations of further bursts from this source should be performed to
confirm this detection, but the results presented here suggest that
radius-expansion bursts from other sources would be prime targets for
{\it RXTE} to search for neutron star spin periods.  A recent report
of 589 Hz pulsations in radius-expansion bursts from an unidentified
source near GRO J1744-28 \markcite{stro1996b} (Strohmayer, Lee, \&
Jahoda 1996b) could be the second example of such a detection, as the
pulsations in those bursts also begin several seconds after the burst
rise is over.  The 363 Hz pulsations from 4U 1728-34, on the other
hand, seem more likely to arise from nuclear flash inhomogeneities.
They appear during the burst rise, and they exhibit frequency
evolution \markcite{stro1996a} (Strohmayer et al.  1996a).  Neither of
these characteristics is displayed by the 524 Hz signal from KS
1731-260.

\acknowledgments

We are grateful to the entire ASM and {\it RXTE} engineering and
science teams for their support. We thank in particular L. Bildsten,
D. Chakrabarty, A. Levine, H. Marshall, R. Remillard, G. Rohrbach, A.
Rots, R.  Shirey and L. Stella for assistance and helpful discussions
about this work. Support for this work was provided in part by NASA
Contract NAS5-30612.

\newpage

\figcaption{Light curve of the burst on 1996 July 14 (UTC) in
three energy bands binned into 0.125-s time bins.  The dashed vertical
lines mark the boundaries of the 2.0-s interval during which the 524
Hz pulsations are observed.}

\figcaption{Power density spectra (PDS) of the event-mode
light curve (2-27 keV) during the peak of the burst in four sequential
1-s intervals as well as the combined 2.0-s interval in which
pulsations are detected.}

\figcaption{Folded pulse profiles for the 2.0-s interval
exhibiting the 523.92 Hz pulsations.  Error bars are $1\sigma$
confidence intervals, calculated from counting statistics.}

\figcaption{Best-fit parameters of a blackbody model to the
2-27 keV spectra, accumulated in 0.25-s intervals: (a) the square root
of the normalization, which is indicative of the blackbody radius at
10 kpc; (b) the temperature; (c) the resulting luminosity at 10 kpc;
(d) the reduced $\chi^{2}$ statistic. Error bars indicate 90\%
confidence intervals; the intervals for the luminosity are smaller
than the dots.  The behavior of these curves demonstrate the
expansion, contraction, and cooling phases of the burst as well as the
Eddington-limited luminosity at $t < 2$ s.  The dashed lines indicate the
region during which the 524 Hz oscillation is detected.}

\newpage

\begin{deluxetable}{ccccc}
\footnotesize
\tablewidth{0pt}
\tablecaption{PCA observations of KS 1731-260}
\tablehead{
\colhead{Obs.} & \colhead{Date} & \colhead{Time on} & \colhead{Time off} &
\colhead{Avg. Rate (c/s)}
}
\startdata
1  & 07/14/96 & 02:25:00 & 03:00:00 & 2807\nl
2  & 07/14/96 & 03:35:00 & 04:36:00 & 2965\nl
3  & 07/14/96 & 05:12:00 & 06:12:00 & 2940\nl
4  & 07/14/96 & 06:48:00 & 06:58:00 & 3036\nl
5  & 08/01/96 & 16:47:00 & 17:46:00 & 1923\nl
6  & 08/01/96 & 18:23:00 & 19:23:00 & 1841\nl
7  & 08/01/96 & 19:59:00 & 20:45:00 & 1899\nl
8  & 08/31/96 & 17:38:00 & 18:06:00 & 2290\nl
9  & 08/31/96 & 18:42:00 & 18:47:00 & 2231\nl
10 & 08/31/96 & 19:16:00 & 19:42:00 & 2215\nl
\enddata
\end{deluxetable}

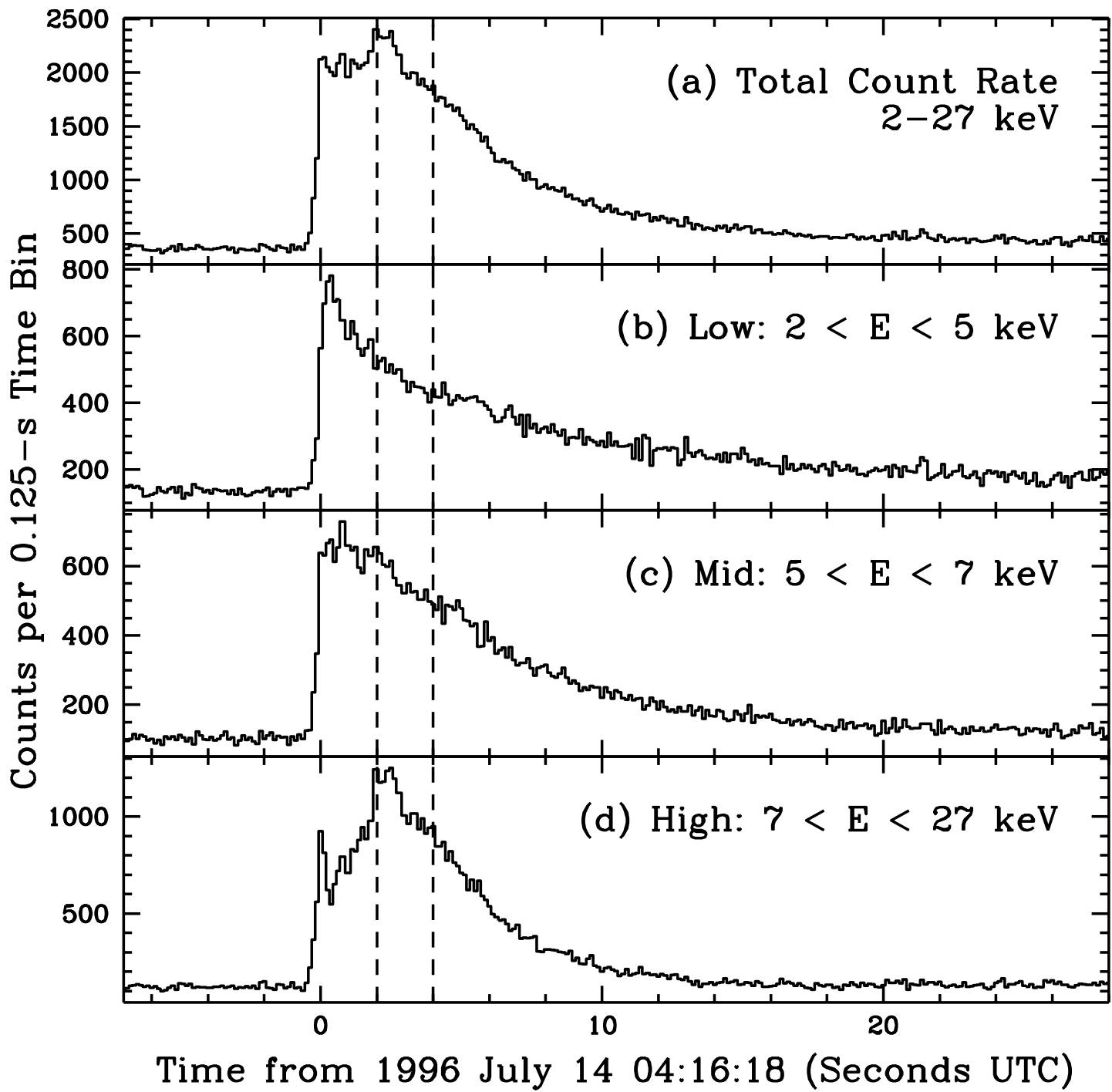
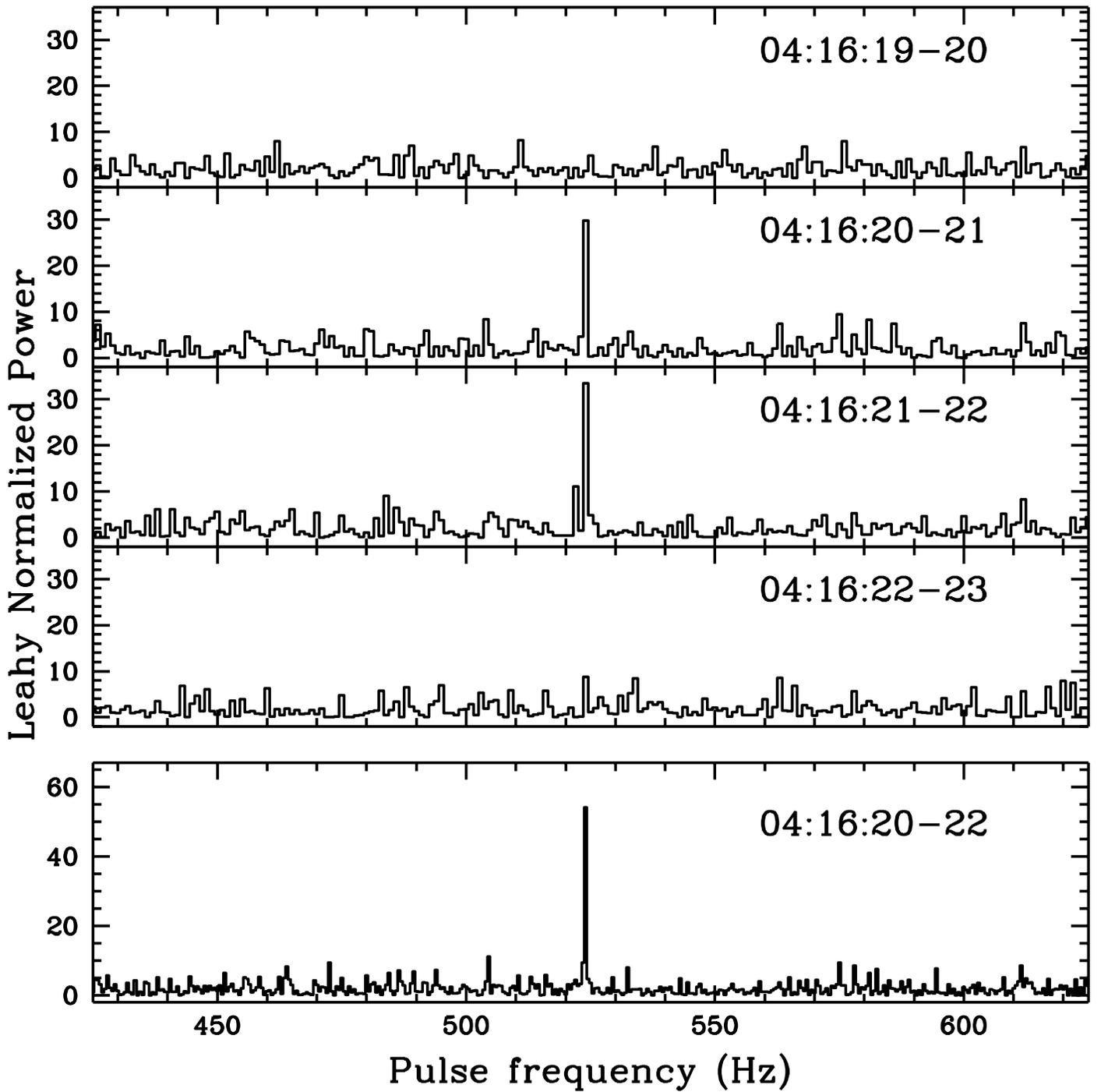
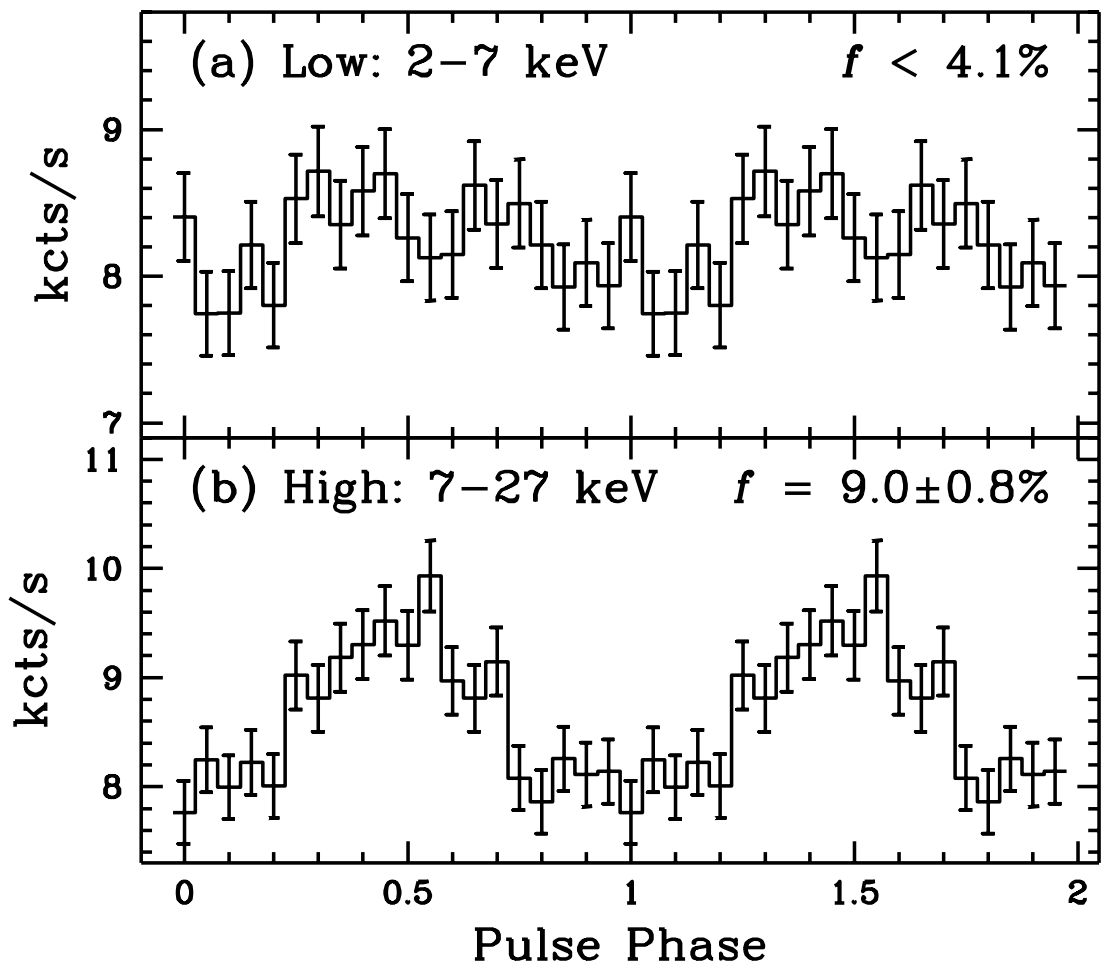
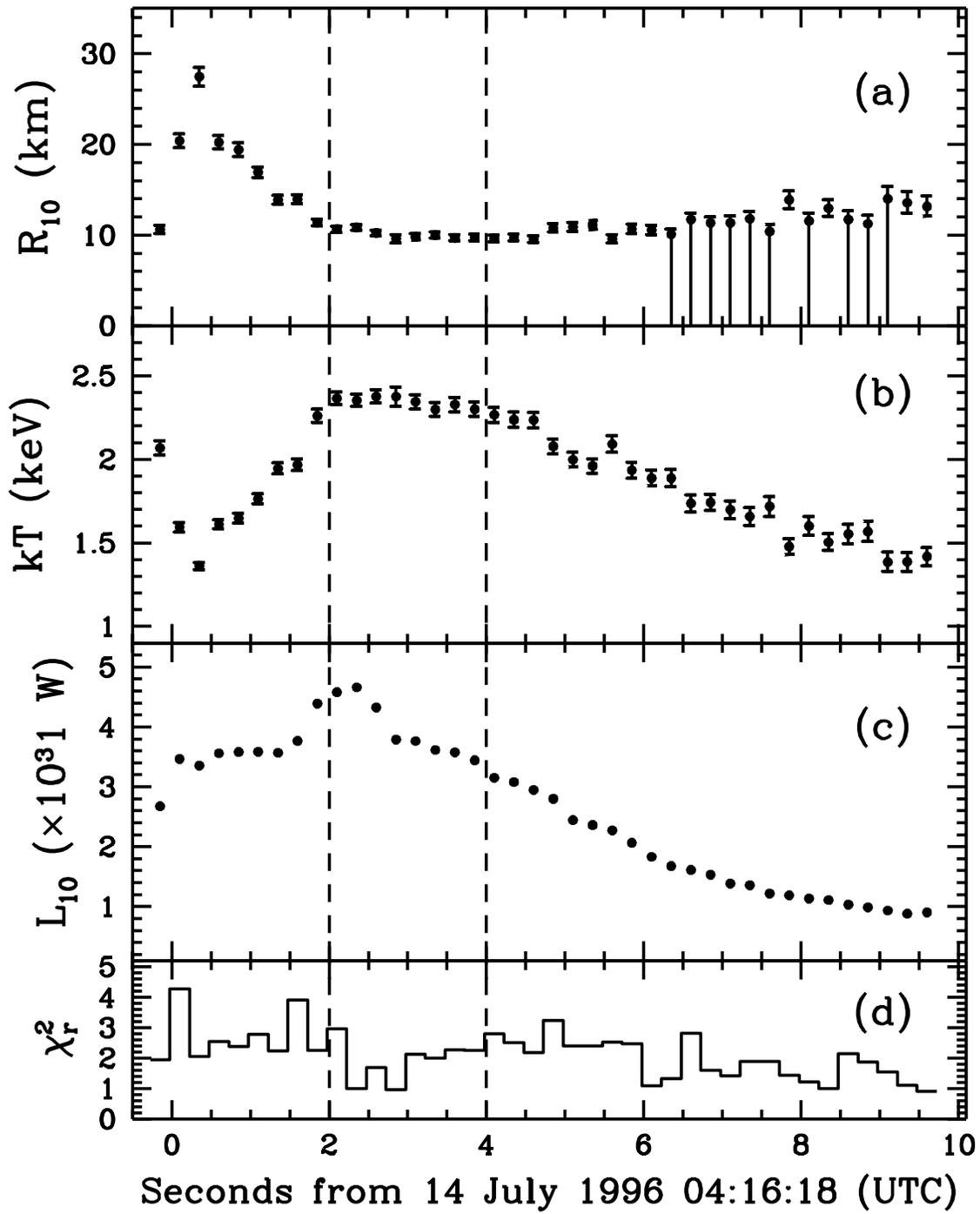
\end{document}